\documentclass[prl,twocolumn,showpacs,preprintnumbers,amsmath,amssymb,revtex4-1]{revtex4-1}
\usepackage{amsfonts, amsmath, bm, bbm, graphicx, epsfig, epstopdf}
\usepackage{dcolumn}
\usepackage{textcomp}
\usepackage{xcolor}
\begin{document}

\title{Polariton Exchange Interactions in Multichannel Optical Networks}

\author{Mohammadsadegh Khazali, Callum R. Murray, and Thomas Pohl}
\affiliation{Department of Physics and Astronomy, Aarhus University , Aarhus 8000, Denmark}

\begin{abstract}
We examine the dynamics of Rydberg polaritons with dipolar interactions that propagate in multiple spatial modes. The dipolar excitation exchange between different Rydberg states mediates an effective exchange between polaritons that enables photons to hop across different spatial channels. Remarkably, the efficiency of this photon exchange process can increase with the channel distance and becomes optimal at a finite rail separation. Based on this mechanism, we design a simple photonic network that realizes a two photon quantum gate with a robust $\pi$-phase, protected by the symmetries of the underlying photon interaction and the geometry of the network. These capabilities expand the scope of Rydberg-EIT towards multidimensional geometries for nonlinear optical networks and explorations of photonic many-body physics.
\end{abstract}

\maketitle

While photons in vacuum lack mutual interactions, the ability to synthetically generate such interactions has high fundamental and technological significance \cite{Chang2014}. Consequently, substantial efforts are currently being directed towards the design of optical interfaces with nonlinearities so large that they can operate at the level of few photons \cite{Birnbaum2005, Schuster2008, Hacker2016, Hwang2009, Maser2016, Tiecke2014, Bhaskar2017, Javadi2015, Snijders2016, DeSantis2017}. Coupling light to cold atomic Rydberg ensembles \cite{Saffman2010} under conditions of electromagntically induced transparency (EIT) \cite{Fleischhauer2005} provides one promising approach to achieving strong, and uniquely long-ranged, photon interactions \cite{Friedler2005, Pritchard2010, Firstenberg2016, Murray2016}. Here, EIT supports the lossless propagation of single photons in the form of so-called Rydberg polaritons\cite{Fleischhauer2002}, while the strong atomic interactions \cite{Lukin2001} prevent the establishment of EIT  and polariton formation for multiple propagating photons. This results in nonlinear phenomena at the level of individual light quanta \cite{Gorshkov2011, Peyronel2012, Firstenberg2013, Baur2014, Gorniaczyk2014, Tiarks2014, Tiarks2016, Tiarks2018, Liang2018}, and, most recently, has made it possible to realize a photon-photon quantum gate using the dispersive nonlinearity induced by Rydberg-state interactions \cite{Tiarks2018}. However, as a direct consequence of the polariton blockade mechanism, the emergent photon interaction has an intrinsic dissipative component \cite{Peyronel2012, Baur2014, Gorniaczyk2014, Tiarks2014, Murray2016a, Murray2018}, and the associated decoherence in this case is detrimental for many perspective applications \cite{Murray2016a, Zeuthen2017, Tiarks2016, Tiarks2018}.

To overcome such obstacles, new ideas are now being explored that go beyond this blockade-induced nonlinearity in Rydberg-EIT systems and allow for coherent effective polariton interactions. This can be achieved through alternative light-matter coupling schemes \cite{Lahad2017, Murray2017}, or by using different Rydberg interactions, as recently demonstrated in \cite{Thompson2017}. By mapping photons onto Rydberg polariton states with dipolar exhange interactions, photons can effectively acquire a long-ranged exchange interaction in which they coherently swap their polariton state during a collision, and acquire a symmetry protected nonlinear phase shift of $\pi/2$ in the process. While this was primarily investigated in a quasi-1D geometry, such polariton exchange interactions in higher dimensions would open up intriguing perspectives for nonlinear photonic networking, in which photons can be effectively coupled across distinct optical modes.

\begin{figure}[t]
\begin{center}
\includegraphics[width=0.9\columnwidth]{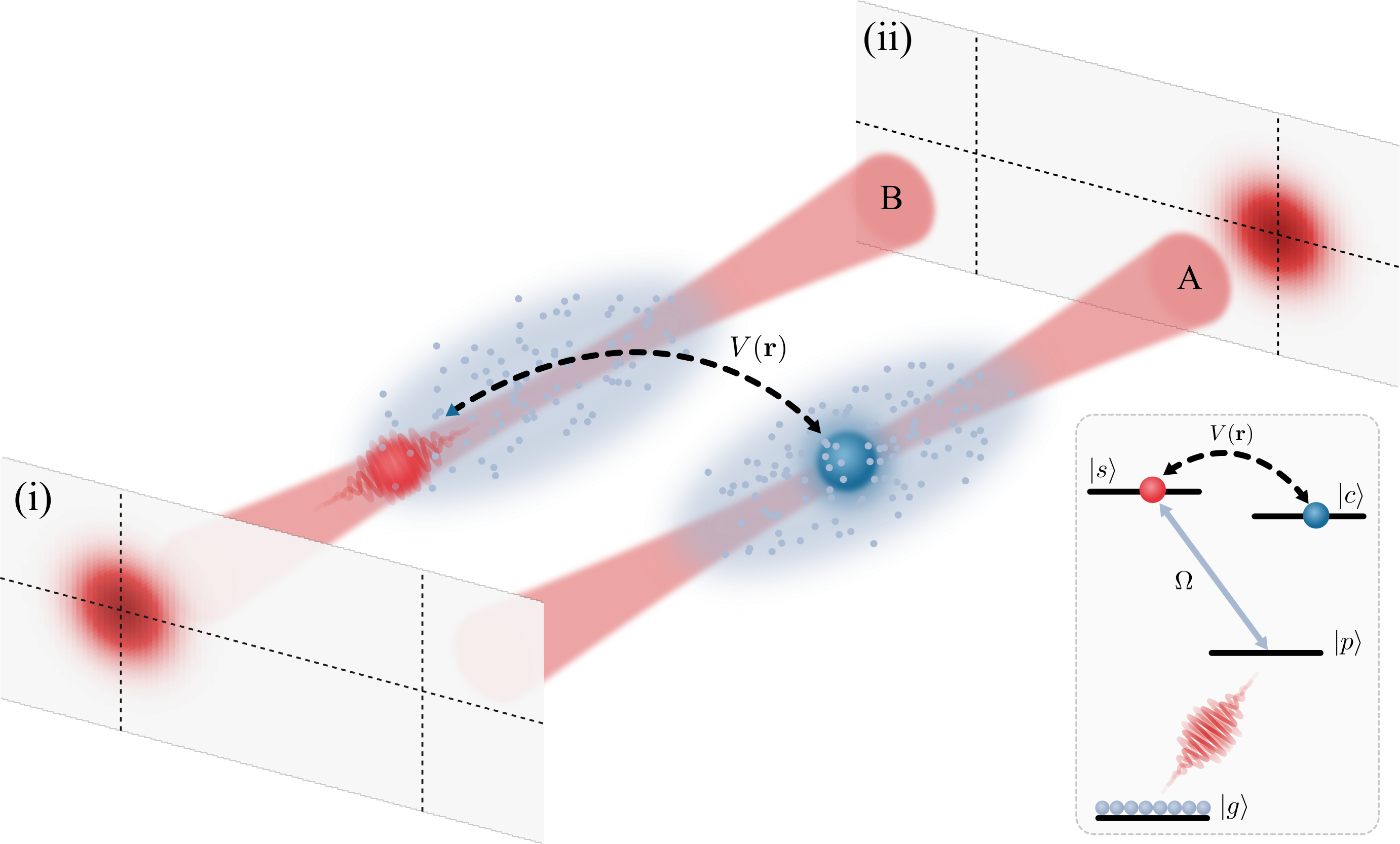}
\end{center}
\caption{\label{fig: 1} Dual-channel setting in which dipolar excitation exchange between atomic Rydberg states mediates an effective photon exchange between two optical channels \textit{A} and \textit{B}. A stored Rydberg spin wave (blue) in channel \textit{A} interacts with the Rydberg-state component of a propagating slow-light polariton (red) in the adjacent channel \textit{B}. Panels (i) and (ii) show the transverse photonic density in each channel before and after the photon collision. The atoms that form the optical medium for both optical modes possess the internal level structure shown in the inset, where $\Omega$ denotes the Rabi frequency for the applied classical field that drives one of the Rydberg-state transitions to establish EIT conditions for the propagating photon.} 
\end{figure}

In this work, we develop the elements of such a nonlinear network by analysing the exchange interaction between polaritons in a dual-channel geometry, as depicted in Fig.\ref{fig: 1}. Surprisingly, we find that the synthetic photon exchange not only remains operational but even becomes more efficient upon increasing the distance between the two spatial channels. This effect arises from the interplay of the coherent dipolar excitation exchange with the directed photon propagation, and yields an optimal rail separation where the exchange efficiency is maximal and exceeds that of one-dimensional head-on collisions. Based on this mechanism we design a simple network that is capable of realizing a symmetry protected $\pi$-phase shift via inherently integrated optical feedback. The resulting gate is shown to feature the most favorable operational fidelity of any Rydberg-atom based photonic $\pi$-phase gate known to date.

The physical setting we consider is illustrated in Fig.\ref{fig: 1}. Here, two transversally separated parallel channels \textit{A} and \textit{B}, formed from tightly focussed free-space optical modes, are incident on an ensemble of atoms in their ground state $|g\rangle$. A single photon propagating in channel \textit{A} is first stored as a stationary spin wave excitation in the Rydberg s state $|s\rangle$. This is achieved via standard light-storage techniques \cite{Fleischhauer2002, Gorshkov2007, Gorshkov2007a, Novikova2007}, using a time-dependent classical control field. Subsequently, a microwave pulse is used to transfer the resulting spin wave excitation to a Rydberg p state $|c\rangle$, which we term the \textit{P} polariton. A second photon in the adjacent optical channel \textit{B} then propagates as a polariton under conditions of EIT with the state $|s\rangle$, which we term the \textit{S} polariton. Here, the photon resonantly couples the $|g\rangle - |p\rangle$ transition with a collectively enhanced coupling strength $G$, while a second classical field continuously drives the $|p\rangle - |s\rangle$ transition with a constant Rabi frequency $\Omega$. The intermediate $|p\rangle$ state decays with a decay rate $\gamma$.
The dipole-dipole interaction $V(r) = C_3/r^3$ between the states $|s\rangle$ and $|c\rangle$ results in a coherent exchange of the excitations at a distance $r$, and thereby induces an effective exchange reaction for the involved polaritons.

The effect of the dipolar interaction on the polariton propagation dynamics can be understood as a competition between polariton blockade and coherent spin exchange, both caused by the dipole-dipole interaction. Specifically, if the \textit{S} and \textit{P} polaritons are within a blockade radius $r_b$, where the interaction $V(r_b)=\Gamma_{\rm EIT}$ starts to exceed the EIT linewidth $\Gamma_{\rm EIT} = \Omega^2/\gamma$ , EIT conditions are violated and the effective polariton interaction becomes purely dissipative \cite{Gorshkov2013}. On the other hand, the hopping radius $r_h$ is the distance over which photon propagation with a group velocity $v_g\approx c \Omega^2/G^2$ takes the same time as the polariton exchange, $V(r_h)^{-1}$. Consequently, it is related to the blockade radius as $r_h=\sqrt{d_b}r_b$ \cite{Thompson2017}, where $d_b$ is the medium's optical depth per blockade radius. Therefore, in the limit of large optical depth, where $r_h>r_b$, the polariton exchange occurs at distances well outside the blockade radius, and the polaritons effectively avoid dissipation due to the polariton blockade. This also implies that van der Waals interactions, previously considered in dual-rail geometries \cite{Maxwell2013,Paredes-Barato2014,Khazali2015,Busche2017}, do not affect the photon dynamics. Overall, this realizes a coherent exchange process, which, in the present dual-channel setting, allows the propagating photon to hop from one spatial channel to another, as illustrated in Fig.\ref{fig: 1}. 

To describe this propagation, we introduce the slowly varying operator $\hat{\mathcal{E}}^{\dagger}(\mathbf{r}, t)$ that creates a photon at position $\mathbf{r}$ and time $t$, which propagates along the optical axis of each channel. Additionally, we introduce the operators $\hat{P}^{\dagger}(\mathbf{r}, t)$, $\hat{S}^{\dagger}(\mathbf{r}, t)$, and $\hat{C}^{\dagger}(\mathbf{r}, t)$ that create collective atomic excitations in the states $|p\rangle$, $|s\rangle$, and $|c\rangle$ respectively. We define $\psi(\mathbf{r}_1, \mathbf{r}_2, t) = \langle 0 | \hat{\mathcal{E}}(\mathbf{r}_1, 0)\hat{C}(\mathbf{r}_2, 0)|\Psi(t)\rangle$ as the amplitude corresponding to a photon at position $\mathbf{r}_1$, and a stored $|c\rangle$ excitation at position $\mathbf{r}_2$, where $|\Psi(t)\rangle$ is the full two-body wave function. As detailed in Ref. \cite{suppl}, this amplitude is governed by the following effective equation in the continuous-wave limit,
\begin{equation}
\label{eq: equation of motion}
\begin{split}
\partial_z \psi(z, \mathbf{r}_\perp, \mathbf{R}_\perp) & =  A(z, r_\perp) \psi(z, \mathbf{r}_\perp, \mathbf{R}_\perp) \\
& + iB(z, r_\perp) \psi(-z, -\mathbf{r}_\perp, \mathbf{R}_\perp),
\end{split}
\end{equation}
where $z$ is the relative separation between the polaritons along the optical axis of the two channels, $\mathbf{r}_\perp$ and $\mathbf{R}_\perp$ are the relative and centre-of-mass coordinates in the plane transverse to the propagation direction, and $r_\perp = |\mathbf{r}_\perp|$. The complex coefficients,
\begin{align}
A(z, r_\perp) & = -\frac{d_b}{z_b} \frac{U^2(z, r_\perp)}{1 + U^2(z, r_\perp)},\label{eq:A}\\
B(z, r_\perp) & = -\frac{d_b}{z_b} \frac{U(z, r_\perp)}{1 + U^2(z, r_\perp)},\label{eq:B}
\end{align}
describe dissipative losses and coherent exchange in Eq.(\ref{eq: equation of motion}), respectively, where $U(z, r_\perp) = V(\sqrt{z^2 + r_\perp^2}) / \Gamma_{\rm EIT}$ denotes the dipole-dipole interaction scaled by the resonant EIT linewidth  $\Gamma_{\rm EIT}$. Denoting the uncorrelated two-body input state as $\psi_{\rm in}(\mathbf{r}_\perp, \mathbf{R}_\perp) \equiv \psi(z\to-\infty, \mathbf{r}_\perp, \mathbf{R}_\perp)$, and the outgoing two-body amplitude as $\psi_{\rm out}(\mathbf{r}_\perp, \mathbf{R}_\perp) \equiv \psi(z\to \infty, \mathbf{r}_\perp, \mathbf{R}_\perp)$, Eq.(\ref{eq: equation of motion}) can be solved to yield, 
\begin{equation}
\begin{split}
\psi_{\rm out}(\mathbf{r}_\perp, \mathbf{R}_\perp) & = T(r_\perp)\psi_{\rm in}(\mathbf{r}_\perp, \mathbf{R}_\perp)  \\
& +  H(r_\perp)\psi_{\rm in}(-\mathbf{r}_\perp, \mathbf{R}_\perp).
\end{split}
\end{equation}
Here, $H(r_\perp)$ and $T(r_\perp)$, respectively, denote the exchange and transmission amplitudes, which yield the probability for the photon to hop to the other spatial mode ($|H(r_\perp)|^2$) or to remain in the initial channel ($|T(r_\perp)|^2$).

\begin{figure}[!h]
\begin{center}
\includegraphics[width=0.95\columnwidth]{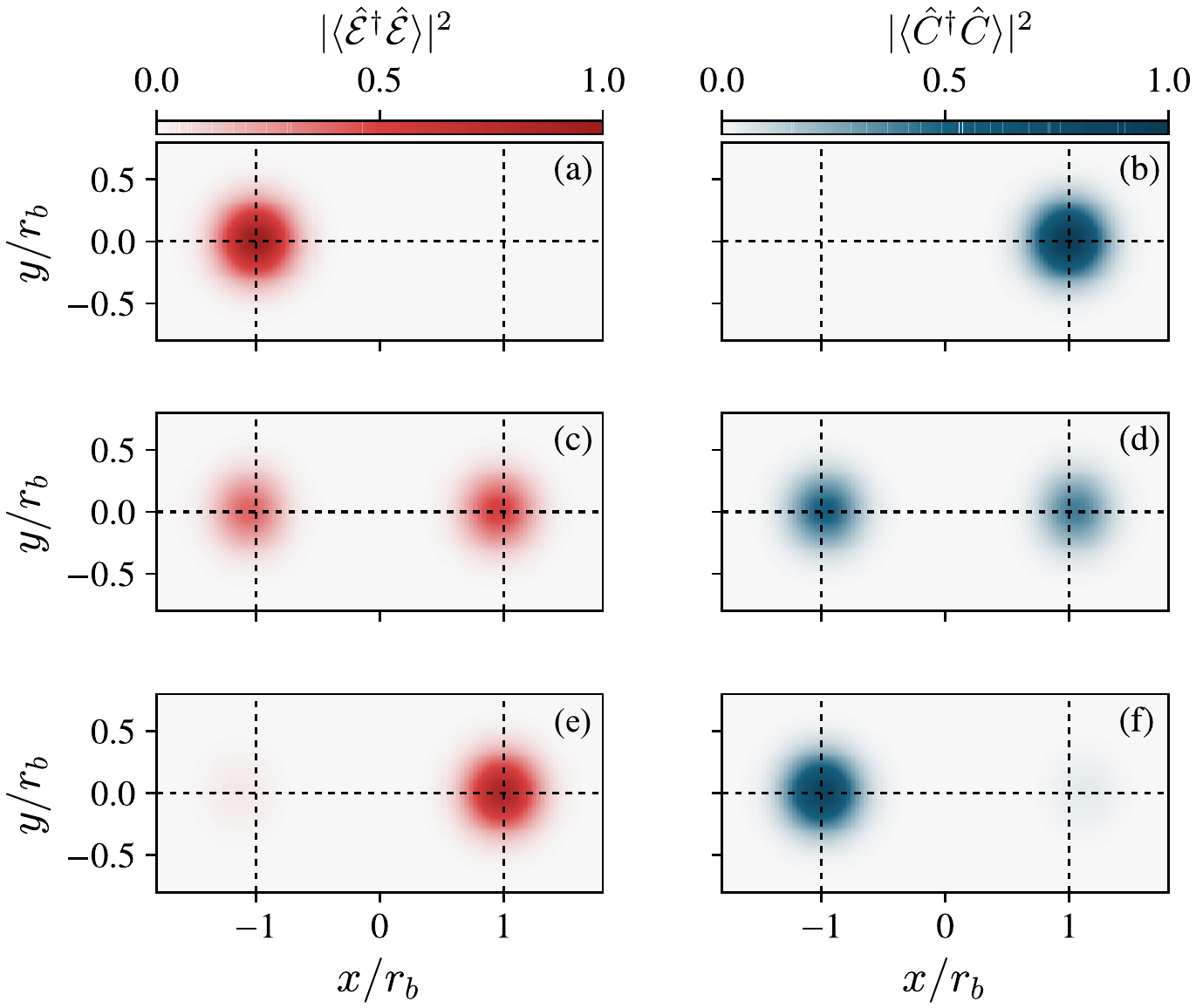}
\end{center}
\caption{\label{fig: 2} (a),(b) Transverse density of incident photon (red) and initial spin wave density of the prepared \textit{P} polariton (blue), for a beam waist of 0.2$r_b$ for each mode, and a channel separation of 2$r_b$. The intersections of the dashed lines indicate the centers of channels \textit{A} and \textit{B} on the right and left, respectively. The transmitted photon density and remaining spin wave density are, respectively, shown in (c),(d) for a blockaded optical depth of $d_b=2$, and in (e),(f) for $d_b=5$.}
\end{figure}

From the solution to $\psi_{\rm out}(\mathbf{r}_\perp, \mathbf{R}_\perp)$, one can obtain the intensity of the transmitted photon $|\langle \hat{\mathcal{E}}^{\dagger}(\mathbf{r}) \hat{\mathcal{E}}(\mathbf{r})\rangle|^2$ and the remaining density of the stored spin wave $|\langle \hat{C}^{\dagger}(\mathbf{r}) \hat{C}(\mathbf{r})\rangle|^2$. Both are shown in Fig.\ref{fig: 2} for Gaussian input modes of the two optical channels. Evidently, a larger optical depth, $d_b$, yields a higher probability for photon hopping, as indicated by an increasing photon intensity at the output of channel \textit{A}, and a concomitant increase of the remaining spin wave density in channel \textit{B}. Importantly, the overall exchange process does not cause significant distortion of each optical mode.

To examine the influence of the rail separation $L$, we define the exchange efficiency 
\begin{equation}
\label{eq: hopping efficiency}
\eta = \left|  \int d\mathbf{r}_\perp \int d\mathbf{R}_\perp H(|\mathbf{r}_\perp|) \left| \psi_{\rm in}(\mathbf{r}_\perp, \mathbf{R}_\perp) \right|^2 \right|^2,
\end{equation}
as the probability for the polaritons to swap channels during a collision. Figure \ref{fig: 3}(a) displays this quantity for both a vanishing and a finite width of the two optical channels. We note that in the former case, the exchange efficiency reduces to $\eta = |H(L)|^2$, which agrees well with the result for experimentally relevant \cite{Thompson2017} input fields with a beam waist of $0.2r_b$. Remarkably, the exchange efficiency is not optimal for a 1D geometry ($L=0$), but rather benefits from a finite separation between the two channels. The emergence of this counterintuitive behavior can be understood from the effective propagation Eq. (\ref{eq: equation of motion}) along with the photon exchange coefficient $B(z, L)$ in Eq.(\ref{eq:B}). For small relative separations between the two polaritons, this coefficient decreases as $B(z, L)\sim (z^2 + L^2)^{-3/2}$ since the diverging dipole-dipole interaction suppresses the Rydberg $|s\rangle$-state excitation and thereby suppresses excitation exchange. A finite rail separation $L$ prevents polaritons from experiencing such short relative separations during a collision, which can therefore result in an enhancement of the overall exchange efficiency. The competition between the dipole blockade and the dipolar excitation exchange gives rise to an optimal polariton exchange efficiency at a specific channel separation $L_{\rm opt}$, which is shown in Fig.\ref{fig: 3}(b) as a function of the blockaded optical depth.

\begin{figure}[!h]
\begin{center}
\includegraphics[width=\columnwidth]{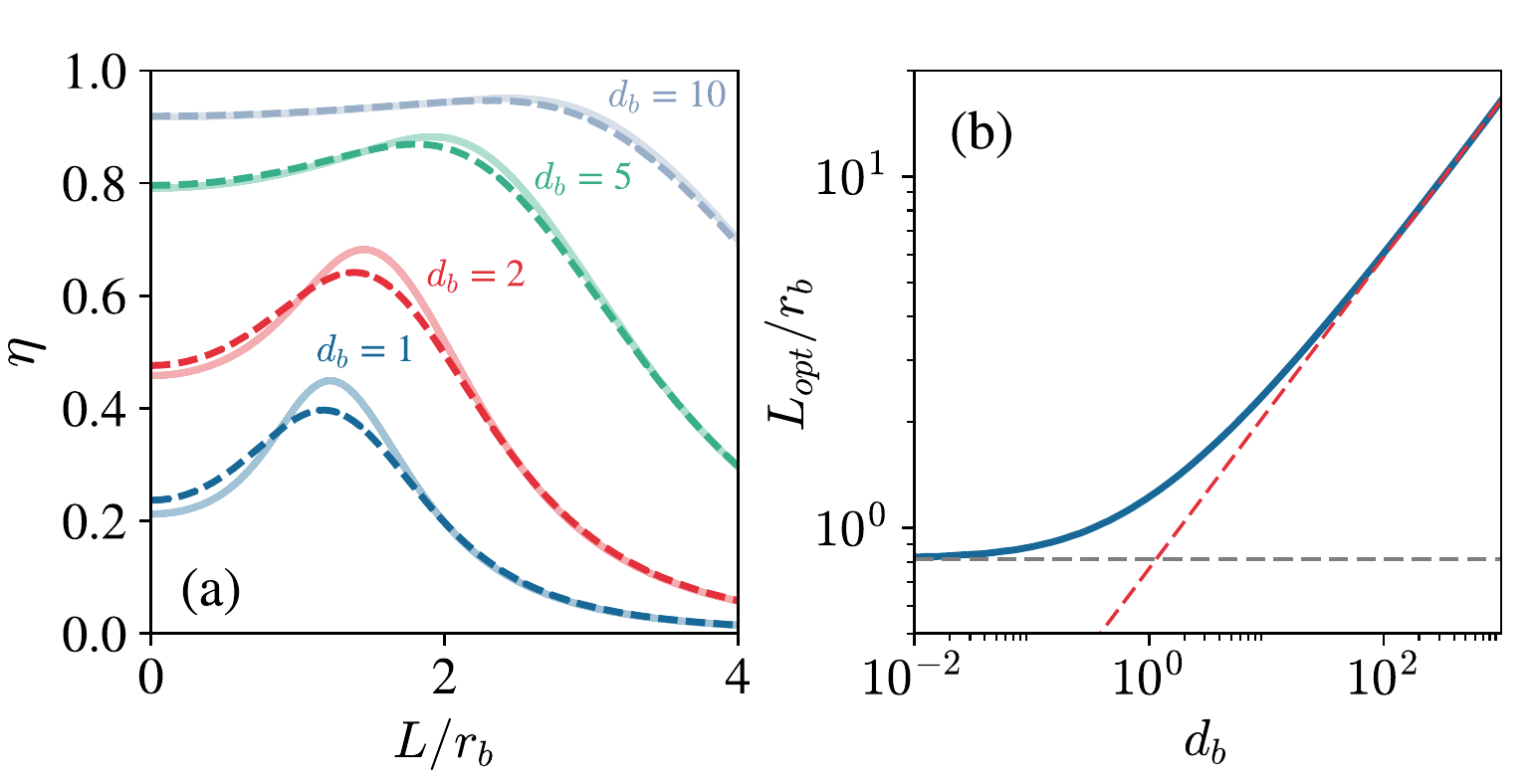}
\end{center}
\caption{\label{fig: 3} (a) Exchange efficiency as a function of the channel separation $L$ for various indicated values of the blockaded optical depth $d_b$. Solid lines show the idealized scenario of zero channel width, while the dashed lines show the result for Gaussian transverse modes with a beam waist of $0.2r_b$. The optimal channel separation as a function of the blockaded optical depth for zero channel width is shown by the solid blue line in panel (b). The horizontal gray dashed line shows the limiting value of $L_{\rm opt}$ for small $d_b$, while the red dashed line shows the power law scaling of $L_{\rm opt} \sim d_b^{0.44}$ in the limit of large $d_b$.}
\end{figure}

For small values of $d_b$, we find that $L_{\rm opt}$ is approximately set by the blockade radius, consistent with the fact that the propagation dynamics are governed by polariton blockade in this regime. We can estimate this limiting value $L_{\rm opt}$ by solving the hopping efficiency to zeroth order in the photon losses as $\eta \approx \tanh^2\left[ \phi(L) \right]$, where $\phi(L) = \int_{-\infty}^{\infty} dz B(z, L)$. This solution is then maximal at the turning point of $\phi(L)$, yielding a numerically determined optimal separation of $L_{\rm opt}\approx0.81 z_b$ [Fig.\ref{fig: 3}(b) gray line]. 

In the large-$d_b$ limit however, we find that the optimal separation obeys a power law scaling $L_{\rm opt} \propto d_b^{\alpha}$, with a numerically determined exponent of $\alpha \approx 0.44$ [Fig.\ref{fig: 3}(b) red line]. This scaling is close to that of the hopping radius $r_h\propto\sqrt{d_b}$, indicating a crossover to polariton exchange dominated dynamics. In this regime, $L_{\rm opt}$ increases with $d_b$ due to the competition between the dissipative and coherent components of the effective polariton interaction. While both the loss coefficient $A(z, L)$ and hopping coefficient $B(z, L)$ increase linearly with $d_b$, the scalings $A(z, L)\sim(z^2 + L^2)^{-3}$ and $B(z, L)\sim(z^2 + L^2)^{-3/2}$ at large transversal separation show that the losses can  be suppressed relative to the hopping by simultaneously increasing $L$.

When the polaritons exchange spin states during a collision in a one-dimensional configuration \cite{Thompson2017}, they acquire a robust phase shift that approaches $\pi/2$ as $d_b\to\infty$, which originates from the symmetries of the underlying effective Hamiltonian. This symmetry-protected phase also emerges in the present situation, whereby the transmitted photon in the output channel A even carries an exact $\pi/2$ phase for any value of $d_b$, since the conditional phase rotation occurs as a direct consequence of the photon hopping. Compared to alternative methods for generating large conditional phase shifts based on polariton blockade \cite{Gorshkov2011, Tiarks2016, Tiarks2018, Murray2017}, this mechanism thus becomes independent of the precise experimental parameters, and is therefore ideally suited to implementing high-fidelity photon-photon phase gate operations. Furthermore, the correlation between the acquired phase and the spatial mode of the outgoing two-photon state offers a natural and convenient way to herald its operation, and is therefore useful for applications in quantum communication \cite{Borregaard2015} and computation \cite{Borregaard2017}. Yet, most of such applications require a phase shift of $\pi$. Achieving this with the current setting therefore implies that the involved polaritons have to interact and exchange spin states exactly twice. 

Indeed, this can be achieved with a simple photonic network of multiple rails. One possible network architecture is depicted in Fig.\ref{fig: 4}. Building on the dual-channel setting shown in Fig.\ref{fig: 1}, this network implements integrated feedback from the output of channel \textit{A} into the input of a third channel \textit{C}, so as to engender a second collision between the involved polaritons. As before, the protocol starts by preparing the stationary \textit{P} polariton in channel \textit{A}, after which the propagating \textit{S} polariton traverses channel \textit{B} [see Fig.\ref{fig: 4}(a.i)]. Following the first successful spin exchange in this case, the \textit{P} polariton is left in channel \textit{B}, while the \textit{S} polariton exits channel \textit{A} and is guided to the input of channel \textit{C} [see Fig.\ref{fig: 4}(a.ii)]. The subsequent second exchange interaction yields the \textit{P} polariton in channel \textit{C} and the \textit{S} polariton back into its original channel \textit{B} [see Fig.\ref{fig: 4}(a.iii)], but with a total acquired two-photon phase of exactly $\pi$. The gate operation is finally completed by retrieving the remaining \textit{P} polariton. While a similar gate operation could be achieved with a closed feedback loop involving only two spatial channels, the open loop configuration shown in Fig.\ref{fig: 4} facilitates the storage and retrieval of the control photon. 

\vspace{3ex}
\begin{figure}[!h]
\begin{center}
\includegraphics[width=\columnwidth]{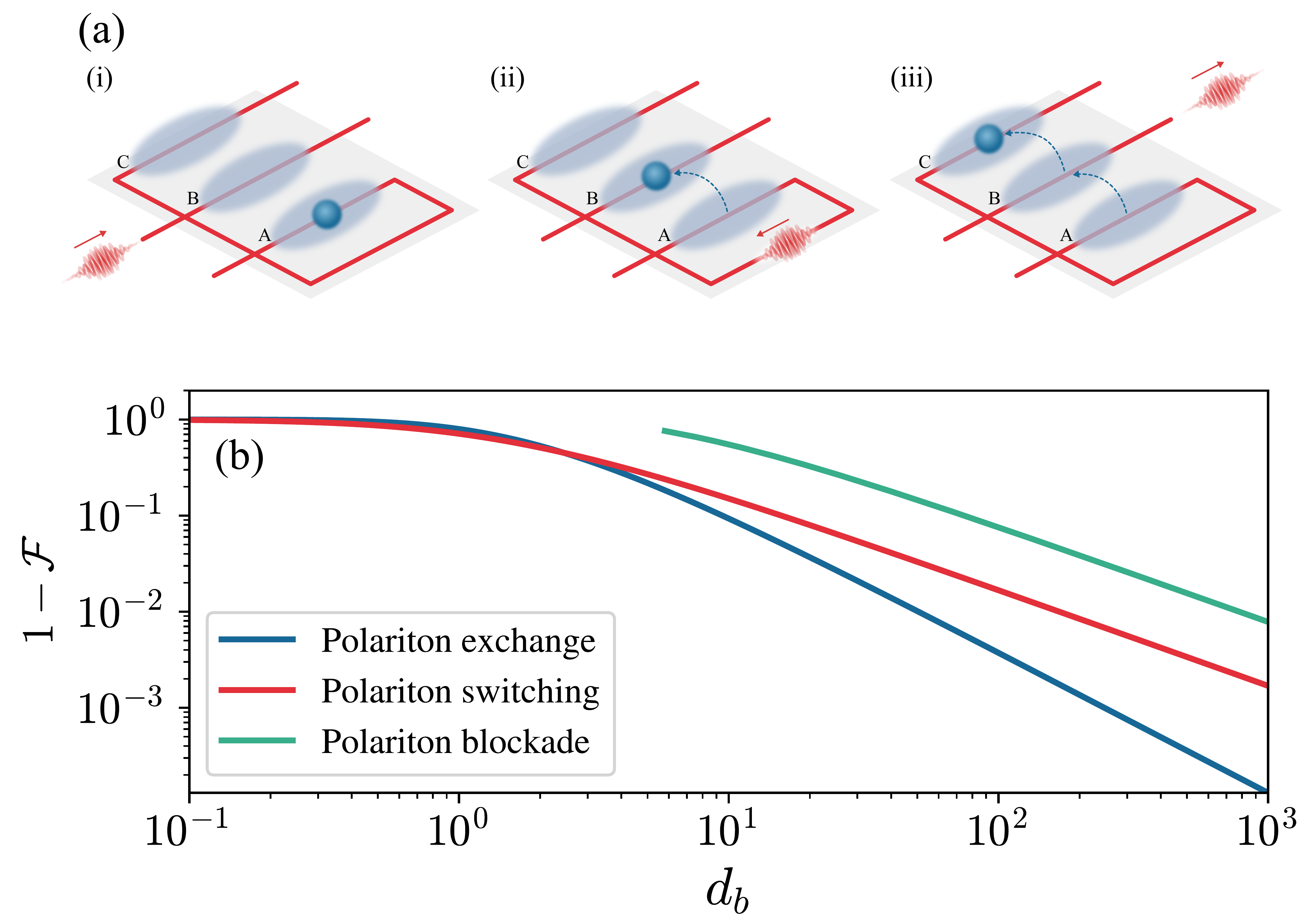}
\caption{\label{fig: 4} (a) Illustration of the optical network for implementing a controlled-\textit{Z} gate via dipolar polariton exchange. In (i) a stationary polariton is initially prepared in channel \textit{A}, while a second propagating polariton traverses channel \textit{B}. (ii) Following a successful exchange, the stationary spin wave is left in channel \textit{B} while the propagating photon is guided into channel \textit{C} via the depicted feedback loop. (iii) Upon entering channel \textit{C} it interacts for a second time with the stationary spin wave in channel \textit{B} and exits the network as shown. (b) $\pi$-phase gate performance for different interaction mechanisms. The blue line shows the gate infidelity of the optimized network shown in (a). The red line shows the corresponding operational infidelity for a gate based on coherent polariton switching \cite{Murray2017}, while the green line reflects the optimal performance for a dispersive $\pi$-phase gate based on the polariton blockade \cite{Gorshkov2011}.}
\end{center}
\end{figure}

To analyze the performance of the described gate scheme, we introduce the following figure of merit,
\begin{equation}
\label{eq: gate efficiency}
\mathcal{F} = \left|  \int d\mathbf{r}_\perp \int d\mathbf{R}_\perp H^2(|\mathbf{r}_\perp|) \left| \psi_{\rm in}(\mathbf{r}_\perp, \mathbf{R}_\perp) \right|^2 \right|^2,
\end{equation}
which characterizes the efficiency for polaritons to exchange spin states exactly twice. Figure \ref{fig: 4}(b) shows the corresponding infidelity $1-\mathcal{F}$ as a function of the blockaded optical depth for an optimal separation $L_{\rm opt}$ between adjacent channels. In the limit of large $d_b$, the gate performance eventually obeys a power-law scaling of $1-\mathcal{F}\propto d_b^{-\alpha}$ with a numerically determined exponent of $\alpha \approx 3/2$, which is the same scaling as the observed photon losses in a 1D geometry \cite{Thompson2017}. 

We remark that the observed $d_b^{-3/2}$ scaling is more favorable compared to alternative gate schemes based on Rydberg blockade with van der Waals interactions \cite{Gorshkov2011, Tiarks2018, Paredes-Barato2014, Khazali2015, Murray2017}. Additionally, a quantitative comparison, see Fig.\ref{fig: 4}(b), shows that the current exchange-mediated gate operation outperforms such existing gate schemes already for moderate values of $d_b$ that are currently available in experiments. While a finite efficiency for storage and retrieval will limit the attainable gate performance, current experiments can achieve quantum memory efficiencies in excess of of $90\%$ \cite{Hsiao2018}. Achieving a similar performance with atomic Rydberg media will require a combination of a high total optical depth in long ensembles, shorter storage times to minimize dephasing effects, and optimization techniques \cite{Gorshkov2007, Gorshkov2007a, Novikova2007}.  We note that because the exchange physics is largely insensitive to the single-photon detuning at large blockaded optical depth, the described gate can operate efficiently with photons of different frequencies \cite{Sun2018}. 

For a complete quantum gate operation \cite{Tiarks2018}, one can encode logical qubits into the left ($|L\rangle$) and right ($|R\rangle$) circular polarization states of the photons. By then implementing ground and Rydberg state EIT with the left and right circularly polarized components, respectively, one can execute a controlled-\textit{Z} gate on the two-photon state. Within this approach, an additional complication arises from the conditioning of the polariton exchange on the presence of two Rydberg polaritons (i.e., two right-circularly polarized photons). Specifically, this generates correlations between the outgoing spatiotemporal mode of the retrieved two-photon state with the two-photon polarization state, due to the time delay caused by slow-light propagation and the pulse advance by $\sim r_h$ during the photon exchange. Moreover, the exchange process for two right-circularly polarized photons moves the remaining \textit{P} polariton to the channel \textit{C}, while all other two-photon states will leave the \textit{P} polariton in the initial channel \textit{A}. While the \textit{P} polariton will eventually be retrieved from channel \textit{C} in all cases, this also induces a time delay conditioned on the two-photon polarization state. Under typical conditions, this delay time can be on the order of $100$ ns, which is similar to the pulse lengths of the involved photonic qubits. However, by synchronizing the delayed two-photon states in a subsequent gradient echo quantum memory \cite{Hosseini2011, Hosseini2012, Sparkes2013, Cho2016}, where photon storage is independent of the arrival time, delay times of up to $10~\mu$s can be corrected with efficiencies of nearly $90\%$ \cite{Cho2016}. 

In conclusion, we have described a strategy for realizing nonlinear quantum optical networks via dipolar exchange reactions between interacting Rydberg polaritons. Considering photon propagation in a dual-channel configuration, we have shown that such a dipolar exchange of atomic states leads to a highly efficient photon exchange at an optimal rail separation, which outperforms the equivalent process in a one-dimensional \cite{Thompson2017} head-on collision between the photons. Based on this capability, we have devised and optimized a simple photonic network that realizes an efficient symmetry-protected controlled-\textit{Z} quantum gate based on inherently integrated nonlinear optical feedback. While we have focussed on the simplest application of the considered setup, the general setting of dipolar photon exchange in multiple dimensions substantially expands the capabilities of Rydberg-EIT and suggests a versatile and powerful framework for more complex quantum networks. Extending the current work to multiple simultaneously incident photons could then make it possible to generate useful $N$-photon entangled states and to explore photonic many-body phenomena.

We thank Aditya Venkatramani, Sergio Cantu, Vladan Vuleti\'{c}, and Mikhail Lukin for helpful discussions. This work was supported by the EU through the H2020-FETOPEN Grant No. 800942 640378 (ErBeStA), by the DFG through the SPP 1929 and by the DNRF through a Niels Bohr Professorship to T.P.
\bibliographystyle{apsrev4-1}
\bibliography{references_hopping}

\newpage ~
\newpage

\clearpage

\title{Supplemental Material for \\ ``Quantum Optical Networks via Polariton Exchange Interactions''}
\author{M. Khazali, C. R. Murray, T. Pohl}
\address{Department of Physics and Astronomy, Aarhus University, Ny Munkegade 120, DK 8000 Aarhus C, Denmark}

\maketitle

\setcounter{equation}{0}
\setcounter{figure}{0}
\setcounter{table}{0}
\setcounter{page}{1}
\makeatletter
\renewcommand{\theequation}{S\arabic{equation}}
\renewcommand{\thefigure}{S\arabic{figure}}
\renewcommand{\thesection}{S\arabic{section}}
\renewcommand{\bibnumfmt}[1]{[S#1]}
\renewcommand{\citenumfont}[1]{S#1}

\onecolumngrid
\vspace{-5ex}

\begin{center}
\large{\bf{Supplemental Material for \\ ``Quantum Optical Networks via Polariton Exchange Interactions''}} \\
\normalsize{M. Khazali, C. R. Murray and T. Pohl} \\
\small{\textit{Department of Physics and Astronomy, Aarhus University , Aarhus 8000, Denmark}}
\end{center}

\section{I. Effective equation of motion for two polaritons}
In this section, we will derive the effective equation of motion describing the two-body dynamics appearing in Eq. (1) of the main text. To this end, we first introduce the slowly-varying photonic operator $\hat{\mathcal{E}}^{\dagger}(\mathbf{r}, t)$ that creates a photon at position $\mathbf{r}$ and time $t$. To describe the atomic dynamics, we then introduce the slowly-varying operators $\hat{P}^{\dagger}(\mathbf{r}, t)$, $\hat{S}^{\dagger}(\mathbf{r}, t)$, and $\hat{C}^{\dagger}(\mathbf{r}, t)$ that create collective atomic excitations in the states $|p\rangle$, $|s\rangle$, and $|c\rangle$, respectively. These operators can be described by the following Heisenberg equations of motion, 
\begin{align}
\partial_t \hat{\mathcal{E}}(\mathbf{r}, t) & = - c \partial_z \hat{\mathcal{E}}(\mathbf{r}, t) - i G \hat{P}(\mathbf{r}, t), \\
\partial_t \hat{P}(\mathbf{r}, t) & = -i G \hat{\mathcal{E}}(\mathbf{r}, t) - i\Omega \hat{S}(\mathbf{r}, t) - \gamma \hat{P}(\mathbf{r}, t), \\
\partial_t \hat{S}(\mathbf{r}, t) & = - i\Omega \hat{P}(\mathbf{r}, t) - i \int d\mathbf{r}^{\prime} V(\mathbf{r} - \mathbf{r}^{\prime}) \hat{C}^{\dagger}(\mathbf{r}^{\prime}, t)\hat{C}(\mathbf{r}^{\prime}, t)\hat{S}(\mathbf{r}, t), \\
\partial_t \hat{C}(\mathbf{r}, t) & = - i \int d\mathbf{r}^{\prime} V(\mathbf{r} - \mathbf{r}^{\prime}) \hat{S}^{\dagger}(\mathbf{r}^{\prime}, t)\hat{S}(\mathbf{r}^{\prime}, t)\hat{C}(\mathbf{r}, t).
\end{align}
Here, we have used the slowly-varying amplitude approximation which assumes a purely linear dispersion relation for the photonic field, describing free-space propagation along the positive $z$-axis at the vacuum speed of light $c$. Photons then drive the low-lying $|g\rangle-|p\rangle$ transition with a collectively enhanced coupling $G\equiv g\sqrt{\rho_a}$ (where $g$ is the single atom coupling and $\rho_a$ is the homogeneous atomic density), while $\gamma$ is the decay rate of the excited state $|p\rangle$. An auxiliary classical field achieves EIT for propagating photons by resonantly driving the $|p\rangle-|s\rangle$ transition with a Rabi frequency $\Omega$. The Rydberg states $|s\rangle$ and $|c\rangle$ then interact via a dipolar exchange interaction $V(r) = C_3/r^3$, where $C_3$ is the interaction coefficient and $r$ is the relative separation between the two excitations. 

The dynamics of a single propagating photon interacting with a single stored spin wave can be described in the Schr\"odinger picture by the following two-body wave function,
\begin{equation}
\begin{split}
|\Psi (t) \rangle & = \int d\mathbf{r}_1 \int d\mathbf{r}_2  EC(\mathbf{r}_1, \mathbf{r}_2, t) \hat{\mathcal{E}}^{\dagger}(\mathbf{r}_1, 0) \hat{C}^{\dagger}(\mathbf{r}_2, 0) | 0 \rangle \\
& + \int d\mathbf{r}_1 \int d\mathbf{r}_2  PC(\mathbf{r}_1, \mathbf{r}_2, t) \hat{P}^{\dagger}(\mathbf{r}_1, t) \hat{C}^{\dagger}(\mathbf{r}_2, t) | 0 \rangle \\
& + \int d\mathbf{r}_1 \int d\mathbf{r}_2  SC(\mathbf{r}_1, \mathbf{r}_2, t) \hat{S}^{\dagger}(\mathbf{r}_1, t) \hat{C}^{\dagger}(\mathbf{r}_2, t) | 0 \rangle.
\end{split}
\end{equation}
Here, $EC(\mathbf{r}_1, \mathbf{r}_2, t) = \langle 0 | \hat{\mathcal{E}}(\mathbf{r}_1, 0) \hat{C}(\mathbf{r}_2, 0) |\Psi(t) \rangle$, $PC(\mathbf{r}_1, \mathbf{r}_2, t) = \langle 0 | \hat{P}(\mathbf{r}_1, 0) \hat{C}(\mathbf{r}_2, 0) |\Psi(t) \rangle$ and $SC(\mathbf{r}_1, \mathbf{r}_2, t) = \langle 0 | \hat{S}(\mathbf{r}_1, 0) \hat{C}(\mathbf{r}_2, 0) |\Psi(t) \rangle$ respectively correspond to the probability amplitudes of finding a photon, $|p\rangle$ or $|s\rangle$ excitation at position $\mathbf{r}_1$, with a stored $|c\rangle$ excitation at position $\mathbf{r}_2$. These amplitudes are then governed by the following equations of motion,
\begin{align}
\partial_t EC(\mathbf{r}_1, \mathbf{r}_2, t) & = - c \partial_{z_1} EC(\mathbf{r}_1, \mathbf{r}_2, t) - i G PC(\mathbf{r}_1, \mathbf{r}_2, t), \\
\partial_t PC(\mathbf{r}_1, \mathbf{r}_2, t) & = - i G EC(\mathbf{r}_1, \mathbf{r}_2, t) - i \Omega SC(\mathbf{r}_1, \mathbf{r}_2, t) - \gamma PC(\mathbf{r}_1, \mathbf{r}_2, t), \\
\partial_t SC(\mathbf{r}_1, \mathbf{r}_2, t) & = -i \Omega PC(\mathbf{r}_1, \mathbf{r}_2, t) - i V(|\mathbf{r}_1 - \mathbf{r}_2|) SC(\mathbf{r}_2, \mathbf{r}_1, t).
\end{align}
Defining $z=z_1-z_2$ and $Z=(z_1+z_2)/2$ as the relative and centre of mass coordinates of the two particles along the propagation direction, and $\mathbf{r}_\perp =\mathbf{r}_{\perp, 1}-\mathbf{r}_{\perp, 2}$ and $\mathbf{R}=(\mathbf{r}_{\perp, 1}+\mathbf{r}_{\perp, 2})/2$ as the relative and centre of mass coordinates in the plane perpendicular to the propagation direction, the above equations of motion can be written as,
\begin{align}
\partial_t EC(z, Z, \mathbf{r}_\perp, \mathbf{R}_\perp, t) & = - c \left[\partial_{z} + \frac{1}{2}\partial_{Z}\right] EC(z, Z, \mathbf{r}_\perp, \mathbf{R}_\perp, t) - i G PC(z, Z, \mathbf{r}_\perp, \mathbf{R}_\perp, t), \\
\partial_t PC(z, Z, \mathbf{r}_\perp, \mathbf{R}_\perp, t) & = i G EC(z, Z, \mathbf{r}_\perp, \mathbf{R}_\perp, t) - i \Omega SC(z, Z, \mathbf{r}_\perp, \mathbf{R}_\perp, t) - \gamma PC(z, Z, \mathbf{r}_\perp, \mathbf{R}_\perp, t), \\
\partial_t SC(z, Z, \mathbf{r}_\perp, \mathbf{R}_\perp, t) & = -i \Omega PC(z, Z, \mathbf{r}_\perp, \mathbf{R}_\perp, t) - i V\left(\sqrt{z^2 + r_\perp^2}\right) SC(-z, Z, -\mathbf{r}_\perp, \mathbf{R}_\perp, t),
\end{align}
where $r_\perp = |\mathbf{r}_\perp|$. By Fourier transforming these equations in time ($t\to \omega$) and in the longitudinal centre of mass ($Z \to K$), we can then derive a single propagation equation for the $EC(z, Z, \mathbf{r}, \mathbf{R}, t)$ amplitude in momentum space as,
\begin{equation}
\partial_z \overline{EC}(z, \mathbf{r}_\perp, \mathbf{R}_\perp, K, \omega) = \bar{A}(z, r_\perp, K, \omega) \overline{EC}(z, \mathbf{r}_\perp, \mathbf{R}_\perp, K, \omega) + i \bar{B}(z, r_\perp, K, \omega) \overline{EC}(-z, -\mathbf{r}_\perp, \mathbf{R}_\perp, K, \omega),
\end{equation}
where $\overline{EC}(z, \mathbf{r}_\perp, \mathbf{R}_\perp, K, \omega)$ is the Fourier transform of $EC(z, Z, \mathbf{r}_\perp, \mathbf{R}_\perp, t)$ in $t$ and $Z$, and the complex coefficients $\bar{A}(z, r_\perp, K, \omega)$ and $\bar{B}(z, r_\perp, K, \omega)$ describing dissipative losses and coherent particle exchange are given by,
\begin{align}
\bar{A}(z, r_\perp, K, \omega) & = - i \frac{\omega}{c} -i \frac{K}{2} + i \frac{G^2}{c(\omega - i\gamma)} + i \frac{G^2\Omega^2}{c(\omega - i\gamma)^2} \frac{\omega - \frac{\Omega^2}{\omega - i \gamma}}{\left( \omega - \frac{\Omega^2}{\omega - i \gamma} \right)^2 - V^2(\sqrt{z^2 + r_\perp^2})}, \\
\bar{B}(z, r_\perp, K, \omega) & = - \frac{G^2\Omega^2}{c(\omega - i\gamma)^2} \frac{V(\sqrt{z^2 + r_\perp^2})}{\left( \omega - \frac{\Omega^2}{\omega - i \gamma} \right)^2 - V^2(\sqrt{z^2 + r_\perp^2})}.
\end{align}
Finally, assuming the stored excitation is described by a sufficiently long spatial mode, and that the propagating photon is described by a sufficiently long temporal mode, it is justified to take the continuous wave limit ($\omega = K = 0$) where translational invariance in space and time are assumed. In this limit, the propagation equation for the amplitude $\psi(z, \mathbf{r}, \mathbf{R}) \equiv \overline{EC}(z, \mathbf{r}, \mathbf{R}, 0, 0)$ is given by,
\begin{equation}
\partial_z \psi(z, \mathbf{r}_\perp, \mathbf{R}_\perp) = A(z, r_\perp) \psi(z, \mathbf{r}_\perp, \mathbf{R}_\perp) + i B(z, r_\perp) \psi(-z, -\mathbf{r}_\perp, \mathbf{R}_\perp),
\end{equation}
as presented in Eq. (1) of the main text, where, 
\begin{align}
A(z, r_\perp) & \equiv \bar{A}(z, r_\perp, 0, 0) = - \frac{db}{z_b}\frac{U^2(z, r_\perp))}{1 + U^2(z, r_\perp))},\\
B(z, r_\perp) & \equiv \bar{B}(z, r_\perp, 0, 0) = - \frac{db}{z_b}\frac{U(z, r_\perp))}{1 + U^2(z, r_\perp))}.
\end{align}
Here, we have introduced $d_b = G^2 z_b / c \gamma$ as the optical depth per blockade radius, and the rescaled potential $U(z, r_\perp) = V(\sqrt{z^2 + r_\perp^2}) / \Gamma_{\rm EIT}$, where $\Gamma_{\rm EIT}=\Omega^2/\gamma$ is the EIT bandwidth.

\end{document}